\documentclass[8pt]{elsart}

\usepackage{graphicx}
\usepackage{latexsym}
\usepackage{wrapfig}
\usepackage[rightcaption]{sidecap}
\usepackage{caption2}
\usepackage{amsmath}
\usepackage{mathrsfs}

\begin{document}
\begin{frontmatter}
\title{Quasinormal modes of Rarita-Schwinger field in
Reissner-Nordstr\"{o}m black hole spacetimes}
\author[shao,gscas]{Fu-Wen Shu}
\author[shao,nao,itp,email]{You-Gen Shen}
\address[shao]{Shanghai Astronomical Observatory, Chinese Academy of Sciences,
Shanghai 200030, People's Republic of China}
\address[nao]{National Astronomical Observatories, Chinese Academy of Sciences,
Beijing 100012, People's Republic of China}
\address[itp]{Institute of Theoretical Physics, Chinese Academy of Sciences,
Beijing 100080, People's Republic of China}
\address[gscas]{Graduate School of Chinese Academy of Sciences,
Beijing 100039, People's Republic of China.}
\thanks[email]{e-mail: ygshen@center. shao. ac. cn}

\begin{abstract}
The Newman-Penrose formalism is used to deal with the quasinormal
modes(QNM's) of Rarita-Schwinger perturbations outside a
Reissner-Nordstr\"{o}m black hole. We obtain four kinds of
possible expressions of effective potentials, which are proved to
be of the same spectra of quasinormal mode frequencies. The
quasinormal mode frequencies evaluated by the WKB potential
approximation show that, similar to those for Dirac perturbations,
the real parts
 of the frequencies increase with the charge $Q$
 and decrease with the mode number $n$, while the dampings
 almost keep unchanged
 as the charge increases.\\
\noindent PACS: number(s): 04.70.Dy, 04.70.Bw, 97.60.Lf \\
Keywords: Black hole, QNM's, Rarita-Schwinger field,
Reissner-Nordstr\"{o}m, WKB approximation.
\end{abstract}

\end{frontmatter}
 \hspace*{3.5mm}For several decades, the QNM's of black holes have
 been of great interest both to gravitation theorists and to
 gravitational-wave experimentalists \cite{ZH,HW,VC} due to the remarkable
  fact that QNM's allow us not only to test the stability
 of the event horizon against small perturbations, but also to probe
 the parameters of black hole, such as its mass, electric charge, and
 angular momentum. QNM's are induced by the external perturbations. For instance, if an
 unfortunate astronaut fall into a black hole, the surrounding
 geometry will undergo damped oscillations. They can be
 accurately described in terms of a set of discrete spectrum of
 complex frequencies, whose real parts determine the oscillation
 frequency, and whose imaginary parts determine the damped rate.
 Mathematically, they are defined as solutions of the perturbation
 equations belonging to certain complex characteristic frequencies
 which satisfy the boundary conditions appropriate for purely ingoing
 waves at the event horizon and purely outgoing waves at
 infinity\cite{CH}.\\
  \hspace*{3.5mm}Recent observational results suggest that our
  Universe in large scale is described by Einstein equations with
  a cosmology constant. Motivated by the recent anti-de Sitter
  (AdS) conformal field theory (CFT) correspondence
  conjecture\cite{JM}, much attention has been paid to the QNM's
  in AdS spacetimes\cite{HW,VC}.  The latest studies on quantum gravity show that
 QNM's also play an important role in this realm due to their close relations to
 the Barbero-Immirzi parameter, a factor introduced
by hand in order that loop quantum gravity reproduces correctly
entropy of the black hole\cite{SH,OD,AC,JB}.\\
 \hspace*{3.5mm} A well-known fact is that quasinormal mode (QN) frequencies are
  closely related to the spin of the exterior perturbation fields\cite{SS}. Previous works on
 QNM's in black holes has concentrated on the perturbation for
 scalar, neutrino, electromagnetic and gravitational
 fields\cite{VC}. However, few has been done for the case of
 Rarita-Schwinger fields, which closely relate to supergravity.
 According to the supergravity theory, the
Rarita-Schwinger field acts as a source of torsion and curvature,
the supergravity field equations reduce to Einstein vacuum field
equations when Rarita-Schwinger field vanishes. This can be seen
from the action of supergravity, namely\cite{DZF},
\begin{equation}
\nonumber I=\int d^{4}x (\textit{L}_2+\textit{L}_{3/2}),
\end{equation}
where $\textit{L}_2$ and $\textit{L}_{3/2}$ represent the
Lagrangian of gravitational and Rarita-Schwinger fields,
respectively. We hence expect to obtain some interesting and new
physics by investigating ONM's of Rarita-Schwinger field.\\
\hspace*{3.5mm} As to QNM's, the first step is to obtain the
one-dimensional radial wave-equation. We usually start with
linearized perturbation equations. Two often used ways are
available for obtaining the linearized perturbation equations. One
is a straightforward but usually complicated way that linearize
the Rarita-Schwinger equation directly and deduce a set of partial
differential equations (one can see Ref.\cite{CVV} for details);
The other is provided by the Newman-Penrose (N-P) formalism, which
end up with partial differential equations in $r$ and $\theta$
instead of ordinary differential equations in $r$.
Torres\cite{GFT} have deduced the linearized Rarita-Schwinger
equation in N-P formalism for a type D vacuum background. We start
with the Rarita-Schwinger equation in a curved background
space-time
\begin{equation}
\nabla_{A\dot{D}}\psi^{A}_{B\dot{C}}=\nabla_{B\dot{C}}\psi^{A}_{A\dot{D}}.
\end{equation}
or, in the Newman-Penrose notation, namely, the Teukolsky's master
equations\cite{SAT,SAT1}
\begin{eqnarray}
\nonumber\{[D-2\epsilon+\epsilon^{*}-3\rho-\rho^{*}](\bar{\Delta}-3\gamma+\mu)\\
-[\delta-2\beta-\alpha^{*}-3\tau+\pi^{*}](\bar{\delta}-3\alpha+\pi)
-\Psi_{2}\}\Phi_{3/2}=0,\label{sp1}
 \end{eqnarray}
and
\begin{eqnarray}
\nonumber \{[\bar{\Delta}+2\gamma-\gamma^{*}+3\mu+\mu^{*}](D+3\epsilon-\rho)\\
-[\bar{\delta}+2\alpha+\beta^{*}+3\pi-\tau^{*}](\delta+3\beta-\tau)
-\Psi_{2}\}\Phi_{-3/2}=0.\label{sp2}
 \end{eqnarray}
Here we have introduced a null tetrad
 $(l^\mu,n^\mu,m^\mu,\bar{m}^\mu)$ which satisfies the
 orthogonality relations $l_{\mu}n^{\mu}=-m_{\mu}\bar{m}^{\mu}=1$ and
 $l_{\mu}m^{\mu}=l_{\mu}\bar{m}^{\mu}=n_{\mu}m^{\mu}=n_{\mu}\bar{m}^{\mu}=0$,
 and the metric conditions $g_{\mu\nu}=l_{\mu}n_{\nu}+n_{\mu}l_{\nu}-
 m_{\mu}\bar{m}_{\nu}-\bar{m}_{\mu}m_{\nu}$. According to these
 conditions, we can take the null tetrad as $ l^{\mu}=(e^{-2U(r)},1,0,0)$,
$n^{\mu}=\frac{1}{2} (1,-e^{2U(r)},0,0)$,
$m^{\mu}=\frac{1}{\sqrt{2}r}\left(0,0,1,\frac{i}{\sin\theta}\right)$,
$\bar{m}^{\mu}=\frac{1}{\sqrt{2}r}\left(0,0,1,-\frac{i}{\sin\theta}\right).
$ The corresponding covariant null tetrad is $l_{\mu}=
(1,-e^{-2U(r)},0,0)$, $n_{\mu}=\frac{1}{2} (e^{2U(r)},1,0,0)$ ,
$m_{\mu}=-\frac{r}{\sqrt{2}}\left(0,0,1,i\sin\theta\right)$,
$\bar{m}_{\mu}=-\frac{r}{\sqrt{2}}\left(0,0,1,-i\sin\theta\right)$.
The non-vanishing spin coefficients read
\begin{eqnarray}
\rho=-\frac{1}{r},\quad
\alpha=-\beta=-\frac{\cot\theta}{2\sqrt{2}r},\quad
\mu=-\frac{e^{2U(r)}}{2r},\quad
\gamma=\frac{1}{4}(e^{2U(r)})^{\prime},
\end{eqnarray}
and only one of Weyl tensors is not zero, i.e.,
$\Psi_{2}=-\frac{(e^{2U(r)})^{\prime}}{2r}$, where a prime denotes
the partial differential with respect to $r$.\\
\hspace*{3.5mm}In standard coordinates, the line element for the
Reissner-Nordstr\"{o}m spacetime can be expressed as
\begin{equation}
ds^{2} =-e^{2U(r)}dt^{2}+e^{-2U(r)}dr^{2}+r^{2}\left( {d\theta
^{2} + sin^{2}\theta d\varphi ^{2}} \right),
\end{equation}
with
\begin{equation}
 e^{2U(r)}
 =1-\frac{{2M}}{{r}}+\frac{Q^{2}}{r^{2}},
 \end{equation}
 where $M$ and $Q$ are the mass and charge of the black hole, respectively. \\
\hspace*{3.5mm}The directional derivatives given in
Eqs.(\ref{sp1}) and (\ref{sp2}), when applied as derivatives to
the functions with a $t$- and a $\varphi$-dependence specified in
the form $e^{i(\omega t+m\varphi)}$, become the derivative
operators
\begin{eqnarray}
D=\mathscr{D}_{0},\quad
\bar{\Delta}=-\frac{\Delta}{2r^{2}}\mathscr{D}^{\dag}_{0},\quad
\delta=\frac{1}{\sqrt{2}r}\mathscr{L}^{\dag}_{0},\quad
\bar{\delta}=\frac{1}{\sqrt{2}r}\mathscr{L}_{0},
\end{eqnarray}
where
\begin{eqnarray}
\mathscr{D}_{n}&=&\partial_{r}+\frac{i\omega
r^{2}}{\Delta}+n\cdot\frac{\Delta^{\prime}}{\Delta},\quad
\mathscr{D}^{\dag}_{n}=\partial_{r}-\frac{i\omega
r^{2}}{\Delta}+n\cdot\frac{\Delta^{\prime}}{\Delta},
\nonumber \\
\mathscr{L}_{n}&=&\partial_{\theta}+\frac{m}{\sin\theta}+n
\cot\theta,\quad
\mathscr{L}^{\dag}_{n}=\partial_{\theta}-\frac{m}{\sin\theta}+n
\cot\theta,
\end{eqnarray}
and
\begin{equation}
\Delta=r^{2}-2Mr+Q^{2}.
\end{equation}
 It is obvious that $\mathscr{D}_{n}$ and
$\mathscr{D}^{\dag}_{n}$ are purely radial operators, while
$\mathscr{L}_{n}$ and $\mathscr{L}^{\dag}_{n}$ are purely angular
operators. After some transformations are made, Eqs.(\ref{sp1})
and (\ref{sp2}) can be decoupled as the two pairs of
equations\cite{SS},
\begin{eqnarray}
[\Delta\mathscr{D}_{-1/2}\mathscr{D}^{\dag}_{0}-4i\omega
r]P_{+3/2}&=&\lambda P_{+3/2},\label{sp5}\\
\mathscr{L}^{\dag}_{-1/2}\mathscr{L}_{3/2}A_{+3/2}&=&-\lambda
A_{+3/2},\label{sp6}
\end{eqnarray}
and
\begin{eqnarray}
[\Delta\mathscr{D}^{\dag}_{-1/2}\mathscr{D}_{0}+4i\omega
r]P_{-3/2}&=&\lambda P_{-3/2},\label{sp7}\\
\mathscr{L}_{-1/2}\mathscr{L}^{\dag}_{3/2}A_{-3/2}&=&-\lambda
A_{-3/2},\label{sp8}
\end{eqnarray}
where $\lambda$ is a separation constant. The reason we have not
distinguished the separation constants in Eqs.(\ref{sp6})-
(\ref{sp8}) is that $\lambda$ is a parameter that is to be
determined by the fact that $A_{+3/2}$ should be regular at
$\theta=0$ and $\theta=\pi$, and thus the operator acting on
$A_{-3/2}$ on the left-hand side of Eq.(\ref{sp8}) is the same as
the one on $A_{+3/2}$ in Eq.(\ref{sp6}) if we replace
$\theta$ by $\pi- \theta$. \\
In Reissner-Nordstr\"{o}m black hole, the separation constant can
be determined analytically\cite{SAT1,NP1,NP}
\begin{equation}
\lambda=\begin{cases}(l+3)(l+1)\quad\quad\quad & \text{for} \quad j=l+s,\\
l(l-2)\quad\quad\quad & \text{for}\quad j=l-s,
\end{cases}
\end{equation}
where $l=2,3,4,\cdots$. Note we only consider the case for $j=l+s$
in our following discussions, the case for $j=l-s$ can be easily
obtained in the same way. Since $P_{+3/2}$ and $P_{-3/2}$ satisfy
complex-conjugate equations (\ref{sp5}) and (\ref{sp7}), it will
suffice to consider the equation
(\ref{sp5}) only.\\
\hspace*{3.5mm}By introducing a tortoise coordinate transformation
$dr_{*}=\frac{r^{2}}{\Delta}dr$, and defining
$\Lambda_{\pm}=\frac{d}{dr_{*}}\pm i\omega$, $Y=r^{-2}P_{+3/2}$,
one can rewrite Eq.(\ref{sp5}) in a simplified form
\begin{equation}
\Lambda^{2}Y+\tilde{P}\Lambda_{-}Y-\tilde{Q}Y=0,\label{sp10}
\end{equation}
where
\begin{equation}
\tilde{P}=\frac{d}{dr_{*}}\ln\frac{r^{6}}{\Delta^{3/2}},\quad
\tilde{Q}=\frac{\Delta}{r^{4}}\left[\lambda-(\frac{2\Delta}{r^{2}}
-\frac{\Delta^{\prime}}{r})\right].
\end{equation}
\hspace*{3.5mm}Transformation theory \cite{CH} shows that one can
transform Eq.(\ref{sp10}) to a one-dimensional wave-equation of
the form $\Lambda^{2}Z=VZ$ by introducing some parameters (certain
functions of $r_*$ to be determined) $\xi(r_*)$, $\chi(r_*)$,
$\beta_1(r_*)$, $T_1(r_*)$, and several constants (to be
specified) $\beta_2$, $T_2$, $\kappa$, $\kappa_1$. If we assume
that $Y$ is related to $Z$ in the manner $Y=\xi VZ+T\Lambda_{+}Z$,
and the relations $T=T_{1}(r_{*})+2i\omega,
\beta=\beta_{1}(r_{*})+2i\omega\beta_{2}$, one can obtain a
equation governing $\beta_2$, $\kappa$, $\kappa_1$ derived from
eq.(\ref{sp10})\cite{SS}
\begin{equation}
\frac{\Delta^{3/2}}{r^{6}}(F+\beta_{2})^{2}-\frac{(F+\beta_{2})
F_{,r_*,r_*}}{F-\beta_{2}}+\frac{({F_{,r_*}}^{2}-\kappa_{1}^{2})F}
{(F-\beta_{2})^{2}}=\kappa,\label{sp23}
 \end{equation}
where we have defined $F=\frac{r^{6}\tilde{Q}}{\Delta^{3/2}}$, and
`$,r_*$' denotes the differential with respect to $r_{*}$. A key
step to obtain the expression of potential $V$ is to seek
available $\beta_{2}$, $\kappa$, and $\kappa_{1}$ whose values
satisfy Eq.(\ref{sp23}). Further study shows that $F$ as defined
in the previous text does satisfy Eq.(\ref{sp23}) with the choice
\begin{equation}
\beta_{2}=\pm2Q,\quad \kappa=0, \quad
\kappa_{1}=\pm\lambda\sqrt{1+\lambda}.
\end{equation}
The potential $V$ can then be expressed as
\begin{equation}
V=-\frac{\Delta^{3/2}}{r^{6}}\beta_2-\frac{(F_{,r_*}-\kappa_{1})(\kappa_1
F-\beta_2F_{,r_*})}{(F-\beta_2)(F^2-\beta_2^2)},\label{potential}
\end{equation}
Note that the sign of $\beta_2$ and $\kappa_1$ can be assigned
independently, so there exists four sets of expressions of $V$.
Here we denote them by $V^j (j=1,2,3,4)$. Chandrasekhar pointed
out that there has a closed relation between the solutions
belonging to the different potentials showed in
equation(\ref{potential}) (see Ref.\cite{CH} \S97(d) for details)
\begin{equation}
K^i Z^i=\left\{
\frac{F+\beta_2^i}{F+\beta_2^j}K^j+\left[i\omega\frac{F-\beta_2^j}
{F+\beta_2^j}-\frac{\kappa_1^jF-\beta_2^jF^{\prime}}{F^2-\beta_2^2}\right]
D^{ij}\right\}Z^j+D^{ij}\frac{dZ^j}{dr_*},\label{zre}
\end{equation}
where
\begin{eqnarray}
K&=&-4\omega^2\beta_2+2i\omega \kappa_1,\\
D^{ij}&=&2i\omega(\beta_2^i-\beta_2^j)+(\kappa_1^i-\kappa_1^j)
\frac{F^2-\beta_2^2}{(F-\beta_2^i)(F-\beta_2^j)}
\end{eqnarray}
One can easily prove that the potentials vanish when we let
$r\rightarrow\pm\infty$
\begin{alignat*}{2}
 V^j&\to e^{\frac{r_*}{2M}}, &\quad\quad& \text{as\quad $ r_* \to - \infty$,}\\
 V^j&\to r^{-2}, && \text{as\quad $ r_* \to + \infty$.}
 \end{alignat*}
 A direct consequence of this property is that the
 wave-function has an asymptotically flat behavior for
 $r\rightarrow\pm\infty$, i.e., $Z\rightarrow e^{\mp i\omega
 r_*}$ (this is just the boundary conditions of QNM's).
It has shown that in asymptotically flat spacetimes, solutions
related in the way showed in Eq.(\ref{zre}) yield the same
reflexion and transmission coefficients, and hence possess the
same spectra of QN frequencies\cite{CH}. Moreover, we can easily
obtain the potentials for negative charge by rearranging the order
of $V^j$ for positive ones since $\beta_2$ equals to $\pm2Q$.
Therefore, we shall concentrate
 just on potential with $\beta_2=2Q (Q>0, say)$ and $\kappa_1=\lambda\sqrt{1+\lambda}$ in
 our following works.\\
 \hspace*{3.5mm}We hence know that the radial equation
(\ref{sp5}) can be simplified
  to a one-dimensional wave-equation of the form
\begin{equation}
\frac{d^{2}Z}{dr_{*}^{2}}+\omega^{2}Z=VZ,
\end{equation}
where
\begin{equation}
V=-\frac{2Q\Delta^{3/2}}{r^{6}}-\frac{(F_{,r_*}-\lambda\sqrt{1+\lambda})(\lambda\sqrt{1+\lambda}
F-2QF_{,r_*})}{(F-2Q)(F^2-4Q^2)}.
\end{equation}
Note that we have written $V^j$ as $V$ because we only work with
one case of the potentials.\\
\hspace*{3.5mm}The effective potential $V(r,Q,l)$, which depends
only on the value of $r$ for fixed $Q$ and $l$, has a maximum over
$r\in(r_+,+\infty)$. The location $r_{0}$ of the maximum has to be
evaluated numerically. An interesting phenomenon is that the
position of the potential peak approaches a critical value when
$l\rightarrow\infty$, i.e.,
\begin{equation}
r_{0}(l\rightarrow\infty)\rightarrow3M.
\end{equation}
\hspace*{3.5mm}Obviously, the effective potential relates to the
electric charge of black hole. Figure 1 demonstrates the variation
of the effective potential $V(r,Q,l)$ with respect to charge $Q$
for fixed $l=2$. From this we can see that the peak value of the
effective potential $V$ increases with $Q$, but the location of
the peak decreases with charge. This is quite consistent with the
case
for Dirac perturbation in Reissner-Nordstr\"{o}m black hole spacetimes.\\
\begin{SCfigure}[1][h]\centering
\includegraphics[width=2.9in,height=2.5in]{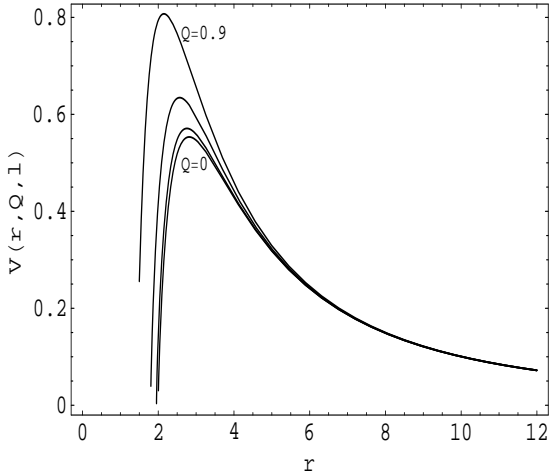}\setcaptionwidth{2.5in}
\caption{Variation of the effective potential for massless
gravitino with $M=1$, $l=2$, $Q=0, 0.3, 0.6, 0.9$ }
\end{SCfigure}
\hspace*{3.5mm}We now evaluate their frequencies by using
third-order WKB potential approximation\cite{SS}, a numerical
method devised by Schutz and Will\cite{SW}, and was extended to
higher orders in\cite{SI1,RA}. Due to its considerable accuracy
for lower-lying modes\cite{SI2}, this analytic method has been
used widely in evaluating QN frequencies of black holes. Noting
that during our evaluating procedures,
  we have let the mass $M$ of the black hole
 as a unit of mass so as to simplify the calculation. The values
 are listed in Table 1, where we only list the values for $l=5$ as an
 example.
 Values for other mode numbers can easily obtained in the same way.
As a reference, we have also evaluated the values (listed in Table
2) for $l=5$ by using the first-order WKB potential
approximation\cite{SW}. Obviously, compared to the first-order
approximation, great improvement, especially
for larger $n$, has been made for third-order approximation.\\
\hspace*{3.5mm} Figure 2 demonstrates the variation of real and
imaginary part of the QN frequencies with different $Q$ and $n$
for $l=5$. It shows that the real part
 of the quasinormal mode frequencies increases with the charge $Q$,
 while decreases with $n$.
 But things are totally different for the imaginary part as showed in the figure,
 whose values almost keep unchanged as  the charge increasing, whereas them increase very
 quickly with the mode number. Furthermore, there is also an
 interesting phenomena that the larger the charge is, the smaller
 effect of $n$ on the real part of QN frequencies may have.
  \begin{table}[t]\centering
\caption{QN frequencies of Rarita-Schwinger field in RN black hole
for $l=5$ (third-order WKB approximation)}
\begin{tabular}{ccccccc}\hline\hline
$Q$ & $n=0$ & $n=1$ & $n=2$ & $n=3$ & $n=4$\\
 \hline
 0   & 1.3273+0.0958i & 1.3196+0.2881i & 1.3048+0.4824i & 1.2839+0.6795i & 1.2582+0.8794i\\
 0.1 & 1.3295+0.0959i & 1.3218+0.2883i & 1.3070+0.4827i & 1.2863+0.6798i & 1.2606+0.8799i\\
 0.2 & 1.3364+0.0960i & 1.3287+0.2888i & 1.3140+0.4835i & 1.2933+0.6809i & 1.2678+0.8813i\\
 0.3 & 1.3481+0.0963i & 1.3405+0.2895i & 1.3260+0.4847i & 1.3055+0.6826i & 1.2802+0.8834i\\
 0.4 & 1.3653+0.0966i & 1.3579+0.2906i & 1.3436+0.4864i & 1.3234+0.6849i & 1.2985+0.8863i\\
 0.5 & 1.3890+0.0970i & 1.3817+0.2918i & 1.3677+0.4884i & 1.3481+0.6876i & 1.3238+0.8896i\\
 0.6 & 1.4206+0.0974i & 1.4136+0.2930i & 1.4001+0.4903i & 1.3811+0.6902i & 1.3576+0.8928i\\
 0.7 & 1.4627+0.0977i & 1.4560+0.2938i & 1.4432+0.4916i & 1.4251+0.6917i & 1.4027+0.8945i\\
 0.8 & 1.5197+0.0976i & 1.5135+0.2932i & 1.5016+0.4904i & 1.4848+0.6898i & 1.4639+0.8916i\\
   \hline\hline
\end{tabular}
\end{table}
\begin{table}[t]\centering
\caption{QN frequencies of Rarita-Schwinger field in RN black hole
for $l=5$ (first-order WKB approximation)}
\begin{tabular}{ccccccc}\hline\hline
$Q$ & $n=0$ & $n=1$ & $n=2$ & $n=3$ & $n=4$\\
 \hline
0   & 1.3358+0.0957i & 1.3618+0.2817i & 1.4077+0.4542i & 1.4657+0.6108i & 1.5300+0.7522i\\
 0.1 & 1.3380+0.0958i & 1.3640+0.2819i & 1.4099+0.4545i & 1.4679+0.6112i & 1.5323+0.7528i\\
 0.2 & 1.3448+0.0959i & 1.3708+0.2824i & 1.4166+0.4554i & 1.4747+0.6125i & 1.5391+0.7546i\\
 0.3 & 1.3566+0.0962i & 1.3825+0.2832i & 1.4282+0.4569i & 1.4862+0.6147i & 1.5507+0.7575i\\
 0.4 & 1.3737+0.0966i & 1.3995+0.2843i & 1.4451+0.4589i & 1.5031+0.6177i & 1.5677+0.7615i\\
 0.5 & 1.3973+0.0970i & 1.4229+0.2857i & 1.4683+0.4614i & 1.5262+0.6215i & 1.5908+0.7666i\\
 0.6 & 1.4288+0.0974i & 1.4541+0.2871i & 1.4991+0.4641i & 1.5567+0.6257i & 1.6213+0.7724i\\
 0.7 & 1.4707+0.0977i & 1.4954+0.2882i & 1.5398+0.4664i & 1.5968+0.6297i & 1.6610+0.7783i\\
 0.8 & 1.5273+0.0975i & 1.5512+0.2880i & 1.5942+0.4671i & 1.6500+0.6318i & 1.7133+0.7823i\\
  \hline\hline
\end{tabular}
\end{table}
\newpage
\begin{figure}[h]\centering
\includegraphics[width=5in,height=3in]{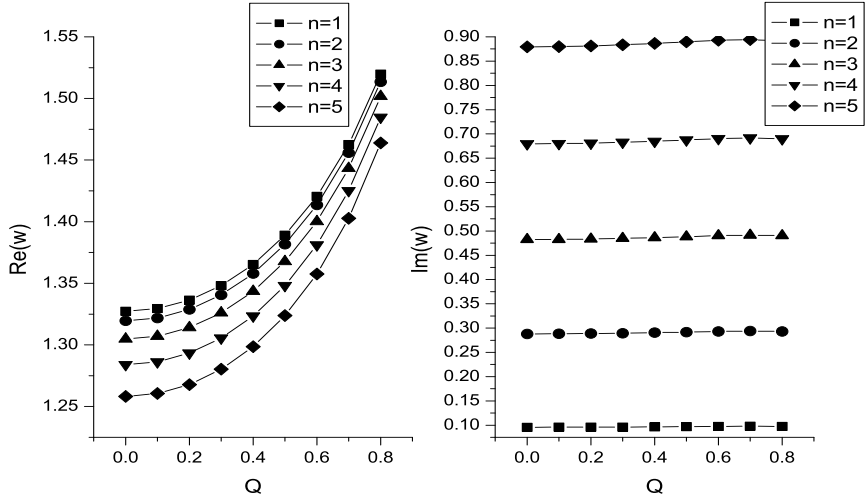}
\caption{Variation of the QN frequencies for $V$ with different
$Q$ and $n$ for $l=5$.}
\end{figure}

\begin{center}\textbf{Acknowledgements}\end{center}
\hspace*{3.5mm}One of the authors(Fu-wen Shu) wishes to thank
Doctor Xian-Hui Ge for his valuable discussion. The work was
supported by the National Natural Science Foundation of China
under Grant No. 10273017.

\end{document}